\documentclass[conference]{IEEEtran}
\IEEEoverridecommandlockouts
\usepackage{cite}
\usepackage{amsmath,amssymb,amsfonts}
\usepackage{algorithmic}
\usepackage{graphicx}
\usepackage{textcomp}
\usepackage{xcolor}
\usepackage{todonotes}
\usepackage{subcaption}
\usepackage[most]{tcolorbox}
\usepackage{xcolor}
\usepackage{multicol}
\usepackage{multirow}
\usepackage{booktabs}
\usepackage{adjustbox}
\usepackage{tikz}
\usepackage{pgfplots}
\usepackage{authblk}
\usepackage{url}
\usepgfplotslibrary{groupplots}
\pgfplotsset{compat=1.18}

\definecolor{colSer}{HTML}{1D4ED8}
\definecolor{colPar}{HTML}{93C5FD}
\definecolor{colLibA}{HTML}{16A34A}
\definecolor{colLibB}{HTML}{86EFAC}

\def\BibTeX{{\rm B\kern-.05em{\sc i\kern-.025em b}\kern-.08em
    T\kern-.1667em\lower.7ex\hbox{E}\kern-.125emX}}
\begin{document}

\title{\huge Cross-Model Cross-Language AI Coding Agent Performance: Accuracy and Speed of Parallel CLRS Algorithms}

 \author[1,3]{Shiqi Cheng}
 \author[1,4]{Evelyne Ringoot}
 \author[1,2,5]{Rabab Alomairy}
\author[1,6]{Alan Edelman}

\affil[1]{Computer Science \& Artificial Intelligence Laboratory,  Massachusetts Institute of Technology, USA.}
\affil[2]{Extreme Computing Research Center, 
Applied Mathematics and Computational Sciences Program, \protect \\
King Abdullah University of Science and Technology, KSA.}
\affil[3]{\quad$^{5}$\textit {sqcheng@mit.edu}
\quad$^{4}$\textit {eringoot@mit.edu}
\quad$^{5}$\textit {rabab.alomairy@mit.edu}
\quad$^{6}$\textit {edelman@mit.edu}
}

\maketitle

\begin{abstract}
AI coding agents have quickly become omnipresent in software engineering. Their serial performance, both in terms of accuracy and speed, has been extensively covered. However, recent initial results suggest their parallel programming capabilities lag behind serial programming capabilities.
This paper presents a cross-language evaluation of three coding agents -- Cursor's Composer 2.0, GPT 5.4, and Claude Sonnet 4.6 -- on parallel code generation across three algorithm categories -- sorting, graph traversal, and search -- in C++, Python, and Julia. For each algorithm and language pair, we prompt a coding agent to produce a parallel implementation from a serial baseline, track the prompting effort required to achieve both functional correctness and performance improvements, and measure speedup against both custom serial baselines and third-party library implementations. We find that coding agents can produce correct parallel implementations with modest prompting effort, but that achieving meaningful speedup is heavily algorithm- and language-dependent. Sonnet 4.6 delivers the strongest overall performance gains, whereas GPT 5.4 produces no measurable speedups despite consistent correctness. C++ is most consistently parallelizable for graph algorithms, while Python and Julia achieve the largest speedups on search algorithms: no single language dominates across all categories. Python and Julia each achieve speedup on some graph algorithms but regress on others. These findings underscore the impact of including runtime performance efficiency as a main LLM performance metric, in addition to accuracy, particularly for parallel implementations.

\end{abstract}


\section{Introduction and related work}

Agentic AI, where coding agents are integrated in the entire software development pipeline, has rapidly shifted the software engineering landscape, leading to extensive research on their relative performance and integration~\cite{wang2025aiagenticprogrammingsurvey}. Many benchmarks have focused on serial code correctness and demonstrated strong accuracy~\cite{chen2021evaluatinglargelanguagemodels,austin2021programsynthesislargelanguage, pmlr-v202-lai23b, 10103177}, in the literature referred to as LLM performance. Other metrics that are classically evaluated include the token cost, security, and code quality~\cite{dong2025survey}. Only more recently have benchmarks arisen that evaluate serial code runtime efficiency performance across a large set of problems~\cite{ICLR2025_06694da0}. 

Many benchmarks have focused on serial algorithms, and evaluations focused on parallel algorithm generation correctness and runtime performance efficiency are more recent, following the publication of several HPC-focused LLMs\cite{chen2023lm4hpc, kadosh2023scopeneedtransformingllms, nichols2023modeling, 10.1007/978-3-031-69577-3_9}. In particular, ParEval\cite{pareval2024} benchmarks of 420 coding tasks spanning 12 computational problem types and six parallel programming models that systematically evaluate coding agents on parallel code generation correctness and performance relative to serial baselines. The study finds coding agents perform significantly worse on parallel code versus sequential code, across all large language models, in particular when targeting MPI along with sparse or unstructured problems. PCEBench~\cite{11078564} similarly compares across computational problems and LLM models.  Further research in this area compares DeepSeek and GPT-4 explicitly, and finds that neither model consistently produces scalable parallel code \cite{deepseek_hpc2025}. Such performance evaluations typically limit themselves to a single programming language -- typically C++ (or occasionally Python~\cite{10.1007/978-3-031-92734-8_5}) --, or to a translation of existing code in C++ or Python to other languages~\cite{10103177}, which does not allow perform in line with natively generated code. 
Similarly, the evaluation of GPU kernel code generation in C++ on massively parallel algorithms and its performance has been a topic of recent research~\cite{ouyang2025kernelbenchllmswriteefficient, tehrani2026finetuninggpt5gpukernel, tariq2025peakperformanceengineeringaiassistant,jaber2026autokernelautonomousgpukernel,wiedemann2026kernelfoundryhardwareawareevolutionarygpu}.
A multi-language approach is used by Valero-Lara et al.\cite{valero2023comparing}, who evaluated Llama-2 and GPT-3, for generating HPC numerical kernels such as AXPY, GEMM, GEMV, across C++, Fortran, Python, and Julia, targeting a range of parallel programming models including OpenMP, OpenACC, CUDA, and Julia's native threading and GPU libraries. They found that correctness correlates with the maturity and public availability of each programming model. However, they do not address speed-up relative to sequential code performance. Critically, their work focuses on algorithms that are massively parallel, such as AXPY, while sorting, graph traversal, and search typically require more reasoning maturity. Where recent research does address AI-optimized performance for serial or parallel algorithms~\cite{11442899,jiang2026hintpilotllmbasedcompilerhint}, it is often found to have a negative effect as the algorithm complexity increases~\cite{rosas2024should, cui2025largelanguagemodelsunderstand}. Indeed, Diehl et al~\cite{10.1007/978-3-031-90200-0_20} perform a cross-language evaluation for a single LLM for complex parallel problems such as stencils and conjugate gradients and find suboptimal performance in both accuracy and parallel speed-up.

In this work, we address the gap of combining performance evaluation with language and AI model evaluation by benchmarking LLMs' capabilities of generating parallel code and combining the approaches proposed by ParEval and Godoy et al on sorting, graph traversal, and search algorithms. The contribution of this paper consists of evaluating Coding Agents-generated parallel code across three languages -- Julia, Python, and C++ -- and across three AI models  -- Cursor's Composer 2.0, GPT 5.4, and Claude Sonnet 4.6 -- on a set of non-massively parallel algorithms drawn from the CLRS benchmark \cite{clrs_text2024}, a classical computer science textbook. We examine how language-specific factors shape the quality and ease of parallel code generation by AI  Coding Agents.


\section{Methods}\label{3}

We select representative algorithms from the CLRS benchmark and evaluate the ability of AI coding agents to parallelize their serial implementations. Experiments are conducted in Python, Julia, and C++ using the Cursor IDE \cite{cursor2024} with Composer 2.0, GPT 5.4, and Claude Sonnet 4.6 as the underlying coding agents. For each algorithm-language pair, the agent is provided with a serial implementation and instructed to generate a parallel version. We record the prompting effort required to achieve functional correctness and evaluate the resulting implementations in terms of correctness and runtime performance.

\subsection{Algorithms}
We evaluate the performance of the AI-generated code on the CLRS benchmark, which provides execution traces of 30 classical algorithms from \textit{Introduction to Algorithms}, and is a standard testbed for neural algorithmic reasoning \cite{clrs_text2024}. We adopt CLRS algorithms in the categories of Sort, Graph Traversal, and Search as the basis for our evaluation, where we task coding agents with generating working parallel implementations of the serial versions of these algorithms.
  For each algorithm and language pair, we prompt a coding agent to produce a parallel implementation starting from a serial baseline, recording the number of semantic and syntactic correctness requirements, and then measure the resulting speedup against the baseline serial implementations as well as third-party libraries. We consider 12 algorithms across three categories, see Table~\ref{tab:algorithms}.

  \begin{table}[t]

\centering

\caption{Algorithms evaluated in this study.}

\label{tab:algorithms}

\begin{tabular}{ll}

\hline

\textbf{Category} & \textbf{Algorithms} \\

\hline

Sorting &

Bubble Sort, Insertion Sort, Merge Sort, \\

& Quick Sort, Selection Sort \\

\hline

Graph &

Breadth-First Search (BFS), Depth-First Search (DFS), \\

& Dijkstra, Bellman--Ford \\

\hline

Search &

Binary Search, Minimum Search, Quickselect \\

\hline

\end{tabular}

\end{table}
\subsection{Iterative approach}
We deploy an iterative approach for code generation. 
In recent literature, iterative or agentic frameworks have been shown to be more effective than single-shot evaluation. For example, the LASSI~\cite{dearing2024lassi} framework incorporates self-correcting feedback loops that respond to compilation and execution results, using targeted prompt engineering and programming language-specific context -- closely related to our iterative prompting protocol. Huang et al.~\cite{huang2025paracoder} similarly proposes a multi-stage agent architecture for parallel code generation and report speedups of approximately $6 \times$ over serial baselines on ParEval tasks. We similarly treat parallel code generation as an iterative process and record the number of interaction rounds required to achieve correctness and a certain level of efficiency.

\subsection{Testing and benchmarking}
All benchmarks were run on a single node with 32 CPU threads (the architecture details can be found in Table \ref{tab:cpu_info}). Benchmarks were timed using language-native timing utilities --- \texttt{std::chrono} in C++, \texttt{time} in Python, and \texttt{BenchmarkTools} in Julia. Timings that are used in this paper are an average of three timed runs after one warm-up run. Additionally, compilation and dataset construction times were excluded from measurements. For each algorithm, the benchmarks have correctness test cases that ensure that the generation doesn't mutate an algorithm into another algorithm. For example, the graph benchmark differentiates between BFS and DFS by checking that the ordering of the node access aligns with the respective algorithms.
\begin{table}[h]
\centering
\begin{tabular}{ll}
\hline
\textbf{Property} & \textbf{Value} \\
\hline
Architecture & x86\_64 \\
Thread(s) per core & 2 \\
Core(s) per socket & 48 \\
Socket(s) & 2 \\
CPU & AMD EPYC 9474F 48-Core Processor \\
CPU Clockspeed (max) & 3600 MHz (4113 MHz)\\
L1d cache & 32K \\
L1i cache & 32K \\
L2 cache & 1024K \\
L3 cache & 32768K \\
\hline
\end{tabular}
\caption{CPU and system information}
\label{tab:cpu_info}
\end{table}

The runtime performance test cases use identical input datasets for serial and parallel timings and across languages, but each of the three runs uses a separate dataset. Input datasets for Sorting and Search benchmarks consist of uniformly random integer arrays with no duplicate values. For Graph benchmarks, graphs are generated with $n$ nodes and $O(n\log n)$ edges. Edge weights (for shortest path algorithms) are uniformly sampled from a fixed range.

\subsection{Prompts}
For each model, the initial prompt was input into the Coding Agents, along with serial implementations of the algorithms and benchmarking code for all three languages. The initial prompt is as follows: 
\begin{tcolorbox}[
    colback=gray!10,
    colframe=gray!10,
    boxrule=0pt,
    arc=3pt
]
Given the serial implementation of each of the five algorithms (bubble, insertion, merge, quick, select) in each language (Julia, Python, C++), provide a parallel multi-threaded implementation of each of the five algorithms. 
\end{tcolorbox}
\begin{tcolorbox}[
    colback=gray!10,
    colframe=gray!10,
    boxrule=0pt,
    arc=3pt
]

Ensure that the parallel implementation adheres to the algorithm itself and do not add constraints that would make the parallel version essentially serial (such as adding a threshold for the size at which something is parallelized). 
\vspace{0.2cm}

 For Julia, use Polyester’s @batch to parallelize the operations. Make sure it is called iteratively and not recursively. For C+++, ensure you are using C++17 and that the data types match this versioning.
\end{tcolorbox}

From there, if correctness is not immediately achieved after the initial prompt, the model is prompted with the error output as follows:  \begin{tcolorbox}[
    colback=gray!10,
    colframe=gray!10,
    boxrule=0pt,
    arc=3pt
]
This is the current benchmarking status of the parallel algorithms:
[Benchmark Output]
\end{tcolorbox}
If the model continues to fail on the correctness tests in the benchmark, human input is given to hypothesize about the potential cause of the error. Finally, runtime performance optimization prompts are given:
\begin{tcolorbox}[
    colback=gray!10,
    colframe=gray!10,
    boxrule=0pt,
    arc=3pt
]
Improve the runtime of the algorithms without changing the algorithms or making the parallel algorithms run in serial. Ensure the continued use \texttt{Polyester.jl} for Julia parallelism.
\end{tcolorbox}
After correctness is achieved, the number of efficiency prompts that are run is based on whether the timings improve. If they do, then the efficiency prompt will be continuously input with the newest benchmarking results. If they don't, the efficiency prompt will be run twice and end. In Cursor, one prompt may internally invoke multiple tool calls/model calls. Thus, there is ambiguity in counting exact model invocations. Therefore, we report the number of user-level prompts rather than raw API calls.




\subsection{Multithreading Packages}
Each language uses different parallelization strategies due to differences in the libraries:
\begin{itemize}
    \item For Python: ProcessPoolExecutor is used. The Global Interpreter Lock (GIL) was a known limitation affecting scalability.
    \item For Julia: Parallelism is implemented using Polyester (as opposed to the better known Base.Threads) because it is faster. However, models frequently produced incorrect or inefficient Polyester usage which required multiple iterations to resolve type errors.
    \item For C++: used \texttt{std::thread} and OpenMP (where applicable).
\end{itemize}

\section{Evaluation}

We evaluate Sorting, Graph, and Search algorithms under the protocol described in Section \ref{3}.

\subsection{Sorting Algorithms}
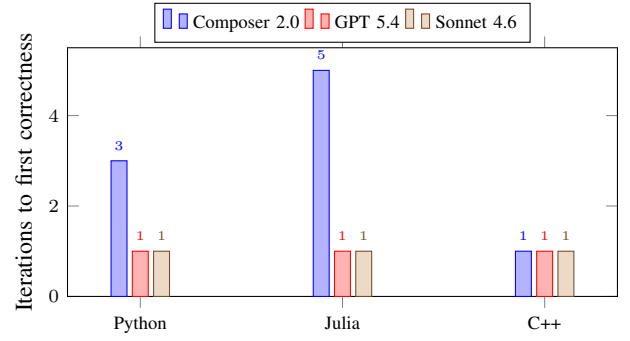
\begin{figure}[t]
  \centering
  \small
  \begin{tikzpicture}
    \begin{axis}[
      width=\linewidth,
      height=0.55\linewidth,
      ybar,
      bar width=6pt,
      ymin=0,
      ylabel={Iterations to first correctness},
      symbolic x coords={Python, Julia, {C++}},
      xtick=data,
      enlarge x limits=0.18,
      legend style={font=\scriptsize, at={(0.5,1.02)}, anchor=south, legend columns=3},
      tick label style={font=\scriptsize},
      label style={font=\small},
      nodes near coords,
      every node near coord/.append style={font=\tiny, yshift=1pt},
    ]
      \addplot coordinates {(Python,3) (Julia,5) ({C++},1)};
      \addplot coordinates {(Python,1) (Julia,1) ({C++},1)};
      \addplot coordinates {(Python,1) (Julia,1) ({C++},1)};
      \legend{Composer 2.0, GPT 5.4, Sonnet 4.6}
    \end{axis}
  \end{tikzpicture}
  \caption{First iteration index at which correctness is achieved on benchmark per language for sorting algorithms.}
  \label{fig:iters-correct}
\end{figure}

In Figure \ref{fig:iters-correct}, the number of iterations to correct parallel code is shown as the number of user prompts before the benchmark successfully runs for all five sorting algorithms per language/coding agent. GPT 5.4 and Sonnet 4.6 reached first correctness in a single prompt for Python, Julia, and C++. Composer 2.0 matched that for C++ but needed three prompts for Python and five for Julia. The Julia gap is consistent with recurring method and type-dimension errors (e.g., MethodError, Int32/Int64 mismatches). Note that GPT 5.4 and Sonnet 4.6 may have reached correctness immediately due to correctness tool calls that the model does within the first user prompt.

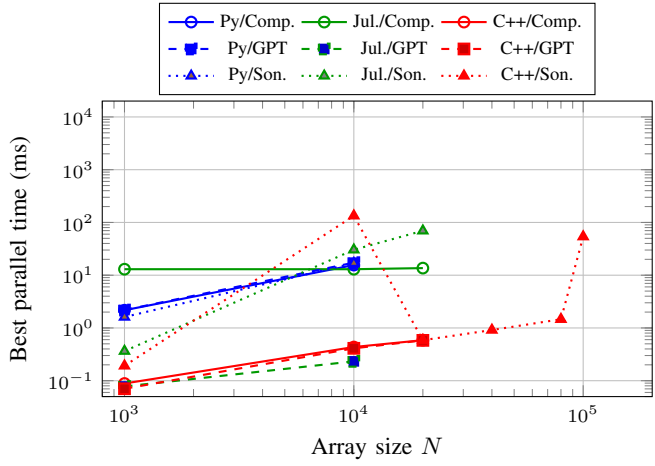
\begin{figure}[t]
  \centering
  \small
  \begin{tikzpicture}
    \begin{axis}[
      width=\linewidth,
      height=0.62\linewidth,
      xmode=log, ymode=log,
      xlabel={Array size $N$},
      ylabel={Best parallel time (ms)},
      xmin=8e2, xmax=2e5,
      ymin=5e-2, ymax=2e4,
      grid=major,
      legend style={
        font=\scriptsize,
        at={(0.5,1.02)}, anchor=south,
        legend columns=3,
      },
      tick label style={font=\scriptsize},
      label style={font=\small},
    ]
            \addplot+[thick, color=blue, mark=o, solid]
        coordinates {(1e3,2.168789) (1e4,15.557284)};
      \addplot+[thick, color=green!60!black, mark=o, solid]
        coordinates {(1e3,12.990782) (1e4,12.986067) (2e4,13.629323)};
      \addplot+[thick, color=red, mark=o, solid]
        coordinates {(1e3,0.088537) (1e4,0.435063) (2e4,0.586744)};

      \addplot+[thick, color=blue, mark=square*, dashed]
        coordinates {(1e3,2.185040) (1e4,17.289596)};
      \addplot+[thick, color=green!60!black, mark=square*, dashed]
        coordinates {(1e3,0.077804) (1e4,0.232784)};
      \addplot+[thick, color=red, mark=square*, dashed]
        coordinates {(1e3,0.070440) (1e4,0.411138) (2e4,0.581570)};

      \addplot+[thick, color=blue, mark=triangle*, dotted]
        coordinates {(1e3,1.613213) (1e4,16.910945)};
      \addplot+[thick, color=green!60!black, mark=triangle*, dotted]
        coordinates {(1e3,0.363833) (1e4,30.333550) (2e4,69.524385)};
      \addplot+[thick, color=red, mark=triangle*, dotted]
        coordinates {(1e3,0.191471) (1e4,133.839900) (2e4,0.586054)
                     (4e4,0.913809) (8e4,1.462514) (1e5,53.797039)};
      \legend{
        Py/Comp., Jul./Comp., C++/Comp.,
        Py/GPT, Jul./GPT, C++/GPT,
        Py/Son., Jul./Son., C++/Son.
      }
    \end{axis}
  \end{tikzpicture}

\caption{Best parallel runtime for each language/coding agent pair at each completed input size for sorting algorithms. Each point is the fastest among the five generated parallel sorting implementations (bubble, insertion, merge, quick, and selection), so the winning algorithm varies across traces: Composer/Julia is fastest with insertion parallel at all completed sizes; Composer/C++ and GPT/C++ use insertion at $N=10^3$ and merge thereafter; GPT/Python uses merge at $N=10^3$ and quick at $N=10^4$; Sonnet/Julia is fastest with bubble parallel; the remaining completed points are fastest with merge parallel.}
  \label{fig:best-parallel}
\end{figure}

Figure~\ref{fig:best-parallel} plots, for each (language, agent) pair, the minimum parallel runtime over the five sorting algorithms at each completed array size $N$. The winning algorithm varies with $N$ and with the language -- for example, Julia/Composer is fastest with parallel insertion at all shown sizes, Julia/Sonnet with parallel bubble, and several C++ traces switch from insertion to merge as~$N$ grows.
Agents therefore do not converge on a single best parallel sort; performance is algorithm- and language-specific. Note that we also observe a spike in best parallel time at $10^4$ for C++/Sonnet. We do not observe the same spike at 16 threads, but display the 32 thread result here for consistency. It is likely that hardware oversubscribing causes this result, due to trade-offs between optimal parallelism and optimal data distribution across threads.




\begin{figure}[!t]
  \centering
  \footnotesize
  \adjustbox{width=\columnwidth,center}{%
    \begin{tikzpicture}
      \begin{axis}[
        width=\columnwidth,
        height=0.65\columnwidth,
        scale only axis,
        ymode=log,
        log basis y={10},
        ymin=0.02,
        ymax=20,
        yticklabel style={font=\footnotesize},
        title style={yshift=2pt, font=\footnotesize, align=center},
        title={%
          Generation speedup at $N=10^4$\\[-0.15em]
          {\scriptsize Marginalizing over Coding Agent (taking best time)}%
        },
        xlabel={Algorithm},
        ylabel={Speedup (serial/parallel)},
        xlabel style={font=\footnotesize},
        ylabel style={font=\footnotesize},
        symbolic x coords={bubble, insertion, merge, quick, selection},
        xtick=data,
        enlarge x limits=0.08,
        x tick label style={
          rotate=35,
          anchor=north east,
          font=\scriptsize,
          inner sep=0.5pt,
        },
        tick label style={font=\footnotesize},
        grid=major,
        minor grid style={densely dashed, thin},
        yminorticks=true,
        clip mode=individual,
        legend style={
          at={(0.02,0.98)},
          anchor=north west,
          legend columns=1,
          font=\scriptsize,
          fill=white,
          fill opacity=0.92,
          draw opacity=1,
          inner sep=2pt,
        },
      ]
        \addplot[dashed, gray!70, thick, forget plot] coordinates {
          (bubble,1) (insertion,1) (merge,1) (quick,1) (selection,1)
        };
        \addplot+[only marks, color=blue, mark=*, thick, mark size=2.5pt,
          mark options={solid, fill=blue!25, draw=blue}] coordinates {
          (bubble,1.06) (insertion,0.24) (merge,1.20) (quick,0.78) (selection,0.41)
        };
        \addlegendentry{Python}
        \addplot+[only marks, color=green!60!black, mark=square*, thick, mark size=3pt,
          mark options={solid, fill=green!15, draw=green!60!black}] coordinates {
          (bubble,1.58) (insertion,0.08) (merge,15.74) (quick,0.81) (selection,1.02)
        };
        \addlegendentry{Julia}
        \addplot+[only marks, color=red, mark=triangle*, thick, mark size=3.5pt,
          mark options={solid, fill=red!15, draw=red}] coordinates {
          (bubble,9.17) (insertion,0.02) (merge,5.27) (quick,0.68) (selection,0.91)
        };
        \addlegendentry{C++}
      \end{axis}
    \end{tikzpicture}%
  }
  \caption{Marginal speedup per language and algorithm at $N=10^4$ for sorting algorithms. A wide variability between the different languages and algorithms is demonstrated: no single language outperforms across all types of algorithms. }
  \label{fig:speedup-10k}
\end{figure}
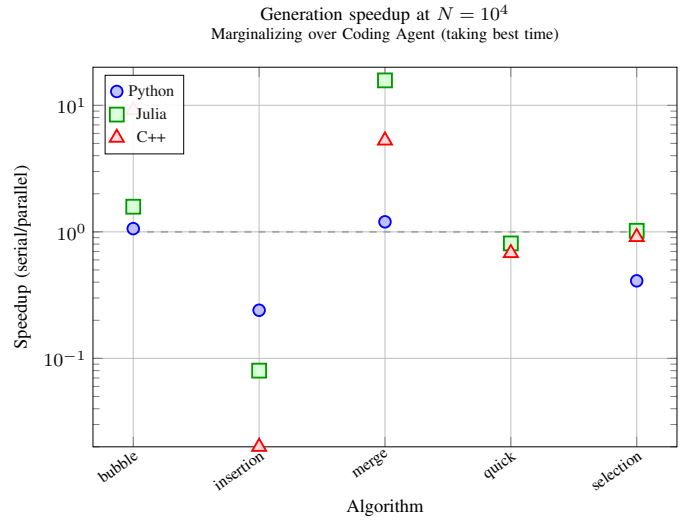

Table \ref{tab:iter-efficiency} records, for the array size $N = 10^4$, how many of the parallel generations are faster than the given serial implementations. GPT reports no wins in any language: it never manages to create a faster parallel than serial implementation, even though correct. Composer achieves $1/5$ faster implementations in Julia (merge algorithms) and $2/5$ in C++ (bubble and merge algorithms), but none in Python. Sonnet is strongest in Julia with faster code for 3 out of 5 algorithms: bubble, merge, and selection. Figure \ref{fig:speedup-10k} (marginal speedup over models) mirrors these results and looks at the five different algorithms. Julia outperforms on merge algorithms; C++ shows strong gains on bubble and merge algorithms; while Python stays near or below the speedup line for most algorithms. Parallel speedups seem to be mostly algorithm-specific, rather than language-specific.

\begin{table}[t]
  \centering
  \caption{Efficiency counter on the number of sorting algorithms with parallel speedup $>1$ (speedup $=\mathrm{serial}/\mathrm{parallel}$) at $N = 10^4$. Number of prompt iterations includes both correctness and efficiency prompts.}
  \label{tab:iter-efficiency}
  \begin{tabular}{@{}llcc@{}}
    \toprule
    Coding Agent & Language & Number of  & Parallel \\
    && iterations & wins \\
    \midrule
    Composer 2.0 & Python & 5 & 0/5 \\
    Composer 2.0 & Julia  & 6 & 1/5 \\
    Composer 2.0 & C++    & 5 & 2/5 \\
    GPT 5.4      & Python & 2 & 0/5 \\
    GPT 5.4      & Julia  & 3 & 0/5 \\
    GPT 5.4      & C++    & 3 & 0/5 \\
    Sonnet 4.6   & Python & 3 & 2/5 \\
    Sonnet 4.6   & Julia  & 3 & 3/5 \\
    Sonnet 4.6   & C++    & 3 & 2/5 \\
    \bottomrule
  \end{tabular}
\end{table}

\subsection{Graph Algorithms}

Table \ref{tab:graph} reports runtimes and speedup ratios at array size $N = 10^5$ for BFS, DFS, Dijkstra, and Bellman-Ford algorithms across all three languages.

C++ achieves parallel speedup across all four graph algorithms and is the only language to do so. Julia achieves performance gains for BFS, DFS, and Bellman-Ford, but shows worse performance for parallel Dijkstra than its serial equivalent, likely because Dijkstra's priority-queue-driven structure introduces synchronization overhead that outweighs the benefits of multithreading. Python obtains only a marginal gain on DFS and BFS.

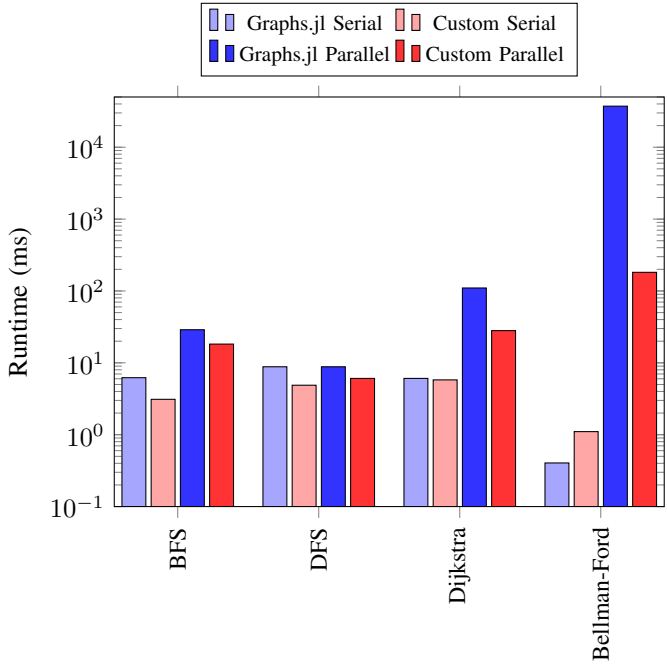
\begin{figure}[t]
\centering
\begin{tikzpicture}
\begin{axis}[
    ybar,
    ymode=log,
    ymin=1e-1,
    ymax=5e4,
    log origin=infty,
    unbounded coords=jump,
    bar width=9pt,
    width=\columnwidth,
    height=7cm,
    ylabel={Runtime (ms)},
    symbolic x coords={BFS, DFS, Dijkstra, Bellman-Ford},
    xtick=data,
    x tick label style={
        rotate=90,
        anchor=east,
        font=\small,
    },
    enlarge x limits=0.15,
    legend style={
        font=\footnotesize,
        at={(0.5,1.05)},
        anchor=south,
        legend columns=2,
        draw=black,
    },
]
\addplot[fill=blue!35] coordinates {
    (BFS,6.212313)
    (DFS,8.816051)
    (Dijkstra,6.073629)
    (Bellman-Ford,0.40441)
};
\addplot[fill=red!35] coordinates {
    (BFS,3.107949)
    (DFS,4.881273)
    (Dijkstra,5.791913)
    (Bellman-Ford,1.103986)
};
\addplot[fill=blue!80] coordinates {
    (BFS,28.878599)
    (DFS,8.816051)
    (Dijkstra,110.022757)
    (Bellman-Ford,37268.470612)
};
\addplot[fill=red!80] coordinates {
    (BFS,18.231233)
    (DFS,6.069929)
    (Dijkstra,28.069488)
    (Bellman-Ford,181.754081)
};
\legend{
    Graphs.jl Serial,
    Custom Serial,
    Graphs.jl Parallel,
    Custom Parallel
}
\end{axis}
\end{tikzpicture}
\caption{Graph algorithm runtimes on a logarithmic scale. Smaller values better.}
\label{fig:graph_bar}
\end{figure}

Figure \ref{fig:graph_bar}, which focuses on Julia, shows graph runtimes on a logarithmic scale, comparing the custom serial and parallel implementations against Graphs.jl serial and Graphs.jl parallel baselines. This figure contextualizes the results against a third-party benchmark: the custom parallel Bellman-Ford implementation competes favorably with Graphs.jl, while the custom Dijkstra implementation falls behind. 

\begin{table}[t]
\caption{Graph Algorithm Runtimes (ms) at $n = 10^5$. Speedup is serial/parallel; values $>1$ indicate parallel is faster.}
\label{tab:graph}
\centering
\begin{tabular}{|l|l|r|r|r|}
\hline
\textbf{Language} & \textbf{Algorithm} & \textbf{Serial (ms)} & \textbf{Parallel (ms)} & \textbf{Speedup} \\
\hline
\multirow{4}{*}{Python}
  & BFS            & 1060.18 & 1063.24 & \textbf{1.00}$\times$ \\
  & DFS            & 1143.16 & 1068.00 & \textbf{1.07}$\times$ \\
  & Dijkstra       & 2142.45 & 2751.64 & 0.78$\times$ \\
  & Bellman-Ford   & 8513.53 & 12730.75 & 0.67$\times$ \\
\hline
\multirow{4}{*}{Julia}
  & BFS            & 110.14  & 109.02  & \textbf{1.01}$\times$ \\
  & DFS            & 440.36  & 427.07  & \textbf{1.03}$\times$ \\
  & Dijkstra       & 332.72  & 537.81  & 0.62$\times$ \\
  & Bellman-Ford   & 627.98  & 494.27  & \textbf{1.27}$\times$ \\
\hline
\multirow{4}{*}{C++}
  & BFS            & 61.10   & 46.08   & \textbf{1.33}$\times$ \\
  & DFS            & 142.39  & 120.34  & \textbf{1.18}$\times$ \\
  & Dijkstra       & 483.70  & 445.38  & \textbf{1.09}$\times$ \\
  & Bellman-Ford   & 786.23  & 535.81  & \textbf{1.47}$\times$ \\
\hline
\end{tabular}
\end{table}

\subsection{Search Algorithms}
The results for Search Algorithms are shown below in Table \ref{tab:search}. At array size $n = 10^6,$ Python achieves most performance benefits on the quickselect algorithm ($1.70\times$), with the minimum algorithm modestly improved ($1.08\times$) and the binary search algorithm slightly slower in parallel ($0.94\times$). This is reasonable as binary search is already $O(\log(n))$ and thread overhead likely dominates. Julia's results are more uneven. The binary search algorithm shows performance gains in its parallel versions; however, the parallel implementations of minimum and quickselect algorithms are dramatically slower, suggesting the \texttt{Polyester.jl}-based parallelizations for these algorithms introduce contention or incorrect usage patterns, a recurring issue identified in the sorting section as well. C++ underperforms across all three search algorithms in parallel. At array size $N = 10^6$, these computations may be too memory-bound for thread-level parallelism to be beneficial.

To support our usage of \texttt{Polyester.jl} for parallelism, we did various experiments between \texttt{Polyester.jl} and \texttt{Base.Threads.jl}, the base mulitthreading library in Julia. Specifically, when comparing the Search algorithms, we can see that at larger sizes especially, \texttt{Polyester.jl} becomes much faster. In Table~\ref{tab:polyester-vs-threads}, we can see that \texttt{Polyester.jl} outperforms \texttt{Base.Threads.jl} for all of the algorithms. These results motivated our choice of \texttt{Polyester.jl} as the multithreading backend for all Julia parallel implementations in this paper.

\begin{table}[t]
  \centering
  \caption{%
    \texttt{Base.Threads} vs.\ \texttt{Polyester.jl} runtimes (ms) for
    parallel search algorithms in Julia at $n = 10^6$. Speedup is
    $\mathrm{Base.Threads} / \mathrm{Polyester}$; values $>1$ indicate
    \texttt{Polyester.jl} is faster.%
  }
  \label{tab:polyester-vs-threads}
  \begin{tabular}{@{}lccc@{}}
    \toprule
    Algorithm & \texttt{Base.Threads} (ms) & \texttt{Polyester} (ms) & Speedup \\
    \midrule
    Binary Search & 1.398 & 0.135 & 10.33$\times$ \\
    Minimum       & 1.316 & 0.242 & 5.44$\times$  \\
    Quickselect   & 40.444 & 37.307 & 1.08$\times$ \\
    \bottomrule
  \end{tabular}
\end{table}

\begin{table}[t]
\caption{Search Algorithm Runtimes (ms) at $n = 10^6$ nodes. Speedup is serial/parallel; values $>1$ indicate parallel is faster.}
\label{tab:search}
\centering
\begin{tabular}{|l|l|r|r|r|}
\hline
\textbf{Language} & \textbf{Algorithm} & \textbf{Serial} & \textbf{Parallel} & \textbf{Speedup} \\
\textbf{} & \textbf{} & \textbf{(ms)} & \textbf{ (ms)} & \textbf{} \\
\hline
\multirow{3}{*}{Python}
  & Binary Search & 29.46  & 31.24  & 0.94$\times$ \\
  & Minimum       & 30.16  & 27.97  & \textbf{1.08}$\times$ \\
  & Quickselect   & 53.56  & 31.53  & \textbf{1.70}$\times$ \\
\hline
\multirow{3}{*}{Julia}
  & Binary Search & 0.020  & 0.015  & \textbf{1.29}$\times$ \\
  & Minimum       & 0.014  & 0.031  & 0.43$\times$ \\
  & Quickselect   & 0.673  & 2.980  & 0.23$\times$ \\
\hline
\multirow{3}{*}{C++}
  & Binary Search & 7.73   & 13.74  & 0.56$\times$ \\
  & Minimum       & 0.098  & 0.114  & 0.86$\times$ \\
  & Quickselect   & 4.90   & 6.62   & 0.74$\times$ \\
\hline
\end{tabular}
\end{table}





We demonstrated, in agreement with prior work, that correctness with minimal prompting is achievable. GPT 5.4 and Sonnet 4.6 consistently reach correctness on the first user prompt for all languages, while Composer 2.0 requires more iterations particularly for Julia, where the strict type system and \texttt{Polyester.jl} usage patterns prove to be recurring failure modes. Comparing LLM models, GPT 5.4 achieves no parallel speedups across all algorithms and languages, despite generating correct implementations. Meanwhile, Sonnet 4.6 is the most effective agent for performance, achieving gains in 7 out of 15 language/algorithm combinations at moderate array sizes of  $N = 10^4$.


\section{Conclusion and Future Work}

This paper evaluated three coding agents—Composer 2.0, GPT 5.4, and Sonnet 4.6—on the task of generating parallel implementations of twelve classical algorithms in Julia, Python, and C++. Consistent with prior studies, we find that modern coding agents are highly effective at producing functionally correct code with minimal user intervention. GPT 5.4 and Sonnet 4.6 achieved correctness on the first prompt across all languages, while Composer 2.0 required additional iterations, particularly for Julia where type constraints and \texttt{\texttt{Polyester.jl}} usage patterns frequently led to errors.

However, our results reveal a substantial gap between correctness and performance. While all agents were generally capable of generating correct parallel implementations, only Sonnet 4.6 consistently translated correctness into runtime improvements, achieving speedups in 7 of 15 language–algorithm combinations. In contrast, GPT 5.4 produced correct implementations but failed to achieve measurable speedups across all evaluated workloads. These findings demonstrate that correctness alone is an insufficient metric for evaluating coding agents on parallel programming tasks. Successful parallelization requires reasoning about workload decomposition, synchronization, memory access patterns, and runtime overheads, all of which remain challenging for current models.

Across languages, C++ delivered the most consistent performance, achieving speedups across all graph algorithm categories and the fastest absolute runtimes in the sorting benchmarks. Julia also demonstrated strong potential, particularly for merge sort and Bellman–Ford, although agent-generated type errors and suboptimal parallelization strategies occasionally limited performance gains. Overall, our study suggests that current coding agents can assist developers in producing correct parallel code, but they are not yet reliable performance-engineering tools.

Future work will expand to investigate larger problem sizes and multicore configurations and evaluate GPU-oriented parallel programming frameworks. An important direction is to study whether performance-aware prompting, automated profiling feedback, or iterative agent-driven optimization can close the gap between correctness and performance. 

\section*{Acknowledgment}
We would like to extend our gratitude for the work and support of members of the Julia Lab. R.A. acknowledges the KAUST Ibn Rushd post-doctoral fellowship. The authors acknowledge the MIT Office of Research Computing and Data, Texas Advanced Computing Center (TACC) at the University of Texas at Austin (TG-CIS250776, allocation  CIS250776 through Advanced Cyberinfrastructure Coordination Ecosystem (ACCESS) program,  supported by U.S. NSF 2138259, 2138286, 2138307, 2137603, and 2138296), and Advanced Micro Devices, Inc. under the AMD University Program’s AI \& HPC Cluster. This material is based upon work supported by the US NSF (CNS-2346520, PHY-2028125, RISE-2425761, DMS-2325184, OAC-2103804, OSI-2029670), DARPA (HR00112490488), DoE (DE-NA0003965), NNSA (DE-NA0004266), USAFR (FA8750-19-2-1000), and DoE NNSA (DE-NA0004266). The U.S. Government, its agencies and employees do not make any warranty, do not assume any liability or responsibility, do not make any endorsement for anything in this report, and do not represent that its use would not infringe privately owned rights. The views and opinions of authors expressed herein are those of the authors alone.

\bibliographystyle{IEEEtran}
\bibliography{references}

\end{document}